\documentclass[10pt,preprint]{aastex}

\newcommand\lsim{\lower0.5ex\hbox{$\; \buildrel < \over \sim \;$}}
\newcommand\gsim{\lower0.5ex\hbox{$\; \buildrel > \over \sim \;$}}
\newcommand\dm{\mbox{$\dot{M}$}}

\newcommand\rout{\mbox{$R_{\rm out}$}}
\newcommand\rg{\mbox{$R_{\rm g}$}}
\newcommand\ms{\mbox{$M_{\odot}$}}

\newcommand\ergs{\mbox{${\rm erg\,s}^{-1}$}}

\shorttitle{A Fallback Disk Model for ULXs} \shortauthors{Li}

\begin{document}

\title{A Fallback Disk Model for Ultraluminous X-ray Sources
}

\author{Xiang-Dong Li}

\affil{Department of Astronomy, Nanjing University, Nanjing
210093, P. R. China}
\email{lixd@nju.edu.cn}

\begin{abstract}
Current stellar evolution models predict that during the core
collapse of massive stars, a considerable amount of the stellar
material will fall back onto the compact, collapsed remnants
(neutron stars or black holes), usually in the form of an
accretion disk. This triggers rapid mass accretion onto, e.g., the
black hole, and produces energetic explosions known as Gamma-ray
bursts. However it is very difficult to prove the existence of an
accretion disk around newborn black holes observationally. Here we
demonstrate that some of the ultraluminous X-ray sources in nearby
galaxies, which are associated with supernova remnants, may be
black holes accreting from their fallback disks, i.e., they have
evolved from collapsars. Since it is almost certain that there is
an accretion disk around these black holes, this would for the
first time lend the observational support to the collapsar model.

\end{abstract}


\section{Introduction}

Gamma-ray bursts (GRBs) are the most brilliant of all astronomical
explosions in our universe. Although we lack direct evidence on
the nature of the central engine, it is widely accepted that GRBs
may involve black holes accreting rapidly through a disk or torus.
In this process some fraction of the accreted mass energy and the
rotational energy of the black holes is extracted and goes into
powering jets that induce GRBs.

There has been accumulating evidence supporting that long-duration
($>2$ s) GRBs are associated with the deaths of massive stars
\citep{g98,f99,b99,g03}. The temporal coincidence of SN2003dh with
GRB030329 \citep{h03} strongly suggests that core collapse events
can give rise to GRBs, thereby favoring the "collapsar" model
\citep{w93,mw99}.

The study of collapsars or "failed supernovae" was initiaed by
\citet{bw83} and the model has been explored as a GRB progenitor
\citep{w93}. During the core collapse of stars with initial mass
higher than $\sim 20-25 \ms$, the black hole may form either
promptly, since a successful outgoing shock fails to be launched
by the collapsed iron core, or in a mild explosion by fallback. In
both cases a significant fraction of the core with mass of
$\sim0.1-5 \ms$ may fall back onto the collapsed remnants over a
period of minutes to hours \citep{w93}. Such a fallback would be
induced by a deep gravitational potential of relatively massive
stars or by a reverse shock propagating inward from the outer
composition interface \citep{c89}. If most of the infalling gas
possesses sufficient angular momentum, the formation of a gaseous
disk is highly feasible.

Although the formation of an accretion disk around a newborn black
hole is a natural outcome in most popular models of GRBs, there
has not been direct observational evidence for the existence of
such a disk. The influence of the disk accretion on the supernova
light curve \citep{bzs00}, and possible gravitational radiation
\citep{m02} have been discussed in the literature.

Theoretically, evolution of the fallback disk around an isolated
compact star has been considered by e.g., \citet{mm89},
\citet{clg90}, \citet{mns93}, and \citet{min97}. The accretion
rates were found to generally have the power-law dependence on
time as $\dm\propto t^{-a}$ with $1<a<1.5$. This is expected by
the self-similar solutions for a viscous disk under the assumption
of constant angular momentum. If disk accretion occurs, it is
natural to expect that the collapsar remnants would eventually
become luminous X-rays sources after the initial, rapidly
accreting phase, for sufficient long time with the disk luminosity
around the Eddington luminosity (see below). In this Letter we
argue that some of the ultraluminous X-ray sources (ULXs)
discovered in nearby galaxies are likely to evolve from
collapsars.

\section{ULXs as black holes accreting from a fallback disk}

ULXs are the most luminous point-like extra-nuclear X-ray sources
found in nearby galaxies, with observed (isotropic) X-ray
luminosities in excess of $10^{39}$ ergs$^{-1}$, roughly
corresponding to the Eddington limit accretion luminosity for a 10
$\ms$ star \citep{f89}. The majority of ULXs are believed to be
accreting black holes, according to their spectral and timing
characteristics \citep{m00,f01}. Currently there are two
categories of models proposed for ULXs. (1) They belong to a new
type of $\sim 10^2-10^5 M_{\odot}$ intermediate-mass black holes
accreting at sub-Eddington rates \citep{cm99,mh02}. (2) They are
traditional stellar-mass black holes in binary systems, but with
either truly super-Eddington X-ray emission \citep{b02}, or
anisotropic \citep{k01}/relativistically beamed emission
\citep{kfm02}. Though some individual ULXs could be
intermediate-mass black holes, the latter model seems to be more
favored by the association of ULXs with star formation regions
\citep{z02}.

Here our interest is focused on the ULXs that were found to be
associated with supernova remnants (SNRs). IC 342 X-1 is one of
the nearest and most comprehensively studied ULXs with an X-ray
luminosity $>10^{40}$ ergs$^{-1}$ in $0.5-10$ keV \citep{m00}.
Recent optical observations \citep{pm02,r03} show a large diameter
($\sim 110$ pc) SNR encircling the position of the ULX, suggesting
that the ULX may be physically related to the SNR. Further
evidence for the association comes from the X-ray ionized nebula
inside the SNR \citep{r03}.

Since the age of the SNR is $\sim 10^5$ years \citep{r03}, the
presence of the ULX implies that it was possibly formed very
recently. If it is in a binary system, mass transfer would occur
in terms of stellar wind capture and/or Roche lobe overflow from
the donor star. However, to produce a luminous X-ray source only
$\sim 10^5$ years after the black hole was formed requires that
both the primary (the black hole progenitor) and the secondary
(donor) stars should have extremely fine-tuned initial masses (see
\citet{bh91} for a discussion on the formation and evolution of
X-ray binaries). A much more plausible explanation is that the
black hole is accreting from the disk originating from the
fallback supernova debris, which implies that this source had
probably experienced a GRB. The inferred initial energy input to
the SNR is indeed at least 2$-$3 times greater than the canonical
energy for an ordinary SNR \citep{r03}. The presence of the ULX
within an unusually energetic SNR strengthens the suggestion that
it may be the aftermath of a GRB.

Another example is the ULX in the SNR MF 16 in NGC 6946
\citep{rc03}, which is possibly $\sim 3500$ years old
\citep{bf94}. If the association is correct, the accreting
material is almost definitely from the fallback disk.

We note that nebulae have been reported to be present at or around
the positions of several ULXs by \citet{pm02}. Some of them could
be SNRs related to the birth of ULXs. These nebulae have kinematic
ages of $\lsim$million years and appear to be directly linked to
the highly energetic formation process of the ULXs or being
inflated by ongoing stellar wind/jet activity. Fallback disk
accretion seems to be a viable mechanism for the X-ray emission
from these sources.

Radio emission was recently discovered from a ULX in the dwarf
irregular galaxy NGC 5408 \citep{k03}. The X-ray, radio and
optical fluxes as well as the X-ray spectral shape are consistent
with beamed relativistic jet emission from an accreting
stellar-mass black hole, which is a strong indicator of disk
accretion. The ULX is found to be close to the star formation
regions of NGC 5408, with a projected displacement of 280 pc, and
the ULX could traverse this distance in $\lsim 3$ Myr with a speed
of 100 kms$^{-1}$ \citep{k03}. This suggests that the source may
belong to young stellar population, i.e., high-mass X-ray binaries
(HMXBs). However, the high X-ray to optical flux ratio $f_{\rm
x}/f_{\rm v}>380$ observed is in sharp conflict with this argument
since Galactic HMXBs have very luminous optical counterpart with
$f_{\rm x}/f_{\rm v}\lsim 1$ \citep{bh91}. Again, this puzzle may
be removed if the black hole is isolated but accreting from the
surrounding disk of the supernova fallback matter. The optical
emission in this case mainly comes from the outer part of the
disk, with a flux much smaller than that in X-ray, as in low-mass
X-ray binaries. A simple estimate from the standard $\alpha$-disk
model \citep{ss73} gives the optical luminosity of the disk to be
\begin{equation}
L_{\rm opt}\simeq 6\times 10^{35}(\frac{M}{3\ms})^{2/3}
(\frac{\dm}{10^{19}\,{\rm gs}^{-1}})^{2/3}\, {\rm ergs}^{-1},
\end{equation}
where $M$ and $\dm$ are the mass of the black hole and the mass
accretion rate, respectively. This value is smaller than the X-ray
luminosity by a factor of $\gsim 1000$ with $\dm\sim 10^{19}\,{\rm
gs}^{-1}$. Actually the optical luminosity may be somewhat higher
than the above value because a part of the hard radiation flux of
the inner disk is reradiated by the X-ray heated, outer disk. But
in this case the influence is mainly limited to infrared
radiation, since X-ray irradiation becomes important at $R>
10^{12}$ cm \citep{s93}.

All of the above evidence points to the fact that some of the ULXs
are likely to be accreting black holes in isolation, or in binary
systems but with no or little mass transfer between the two stars.
The source of the accreting material results from the fallback
matter during the (failed) supernovae that produce the black
holes. From the work of \citet{min97}, we can roughly estimate the
mass accretion rate onto the black hole as,
\begin{equation}
\dm\simeq 2.7\times 10^{19}(\frac{M}{3 \ms})(\frac{\Delta M}{0.5
\ms})(\frac{\alpha}{0.01})^{-1.35}(\frac{t}{10^5\,{\rm
yr}})^{-1.35}\, {\rm gs}^{-1},
\end{equation}
where $\Delta M$ is the amount of fallback material and $\alpha$
the viscosity parameter. It is clearly seen that the black holes
can be ultraluminous, super-Eddington, X-ray sources with an X-ray
lifetime up to $\sim 10^5$ yr for typical values of the
parameters, if their X-ray emission is mildly or relativistically
beamed. The birth rate of hypernovae \citep{p98}, as the
observational consequences of the collapsar model, could be $\sim
10^{-3}$ yr$^{-1}$ in the Galaxy \citep{h99}, compared to a few
$10^{-5}$ yr$^{-1}$ for HMXBs. The ultraluminous X-ray lifetime of
the latter may last for several $10^7$ yr \citep{prh03}. From this
one may expect that a considerable ($\gsim 10\%$) fraction of the
"young" ULXs may produce X-ray emission with the fallback disk
accretion. This provides observational support to the prediction
of the collapsar model.

\section{Discussion and conclusions}

We have argued that fallback disk accretion may play an important
role in producing the X-ray emission of some ULXs that possibly
associated with SNRs or star formation region. This association
rules out mass transfer in binaries as the main energy source,
since there is little time for the donor stars (if they exist) to
evolve and transfer enough mass to the black holes.

The discovery of the optical counterpart to ULX NGC 3031 X-11 is
not in conflict with our argument, which was identified as a 23
$\ms$ O8V star \citep{lbs02}. For this ULX system, since the
companion is a main-sequence star, its stellar wind should not be
so strong to power the X-ray emission, suggesting that the black
hole may be accreting gas via Roche lobe overflow. But both
observations and binary evolution calculations have shown that, it
is much more likely that the secondary is evolved during mass
transfer in luminous HMXBs \citep{bh91,prh03}. In this case
accretion from the fallback disk could be more important than mass
transfer between the two stars.

If some of the ULXs are really related to collapsars (and hence
GRBs), then the post-GRB objects should possess a fallback disk
with a high mass accretion rate, and emit intense high-energy
radiation. However, no such emission has been observed so far from
these objects. The reason is generally attributed to photon
trapping during hypercritical accretion \citep[e.g.][]{min97}. But
more recent slim disk calculations with photon trapping effect
taken into account \citep{o02} show that the disk luminosity could
be kept around the Eddington luminosity, even if the mass
accretion rate is extremely high. Perhaps not all collapsars can
produce GRBs. According to \citet{n01}, the accretion flow in a
putative GRB central engine can be either a convection-dominated
accretion flow (CDAF) if the accreting mass is introduced at a
large outer radius $\rout$, or a neutrino-dominated accretion flow
(NDAF) if $\rout$ is relatively small, i.e., $\rout\lsim
10-100\rg$, where $\rg=2GM/c^2$ is the Schwarzschild radius
\citep[see also][]{km02}. Both types of accretion flows are
radiatively inefficient, but the latter is much more plausible for
a GRB engine since the mass loss during accretion is much smaller
than in the former case. \citet{km02} have studied the properties
of an NDAF with mass of $1 \ms$ and size of $\sim 5\times 10^6$
cm. From their work we can estimate the disk luminosity to be as
low as $10^{31}-10^{32}\, \ergs$. On the other hand, when the disk
evolves to extend to larger radius (or the disk size is initially
relatively large), the accretion may occur as a CDAF \citep{n01}.
It is  not yet clear how bright the CDAF is. But its radiation
luminosity should also be low because the viscous heating rate is
very small in a CDAF since there is no shear stress \citep{n00}.
So in both cases we may expect that the collapsar remnants are dim
sources at their early stage. Full multidimensional MHD
simulations coupled with radiation hydrodynamics are necessary to
settle this issue.

ULXs were also found in elliptical galaxies \citep{sib01,c03}. It
has been suggested that these ULXs may be soft X-ray transients,
i.e., wide low-mass X-ray binaries, at their high (outburst) state
\citep{k02}. Recent analysis by \citet{c03} shows that more than
20\% of all ULXs found in spirals originate from the older (pop
II) stellar populations, indicating that many of the ULXs that
have been found in spirals galaxies are in fact pop II ULXs, like
those in elliptical galaxies. This would require a fairly large
birth rate of ULXs (see \citet{k01} for a discussion), which is
not easily accounted for by wide low-mass X-ray binaries with an
estimated birth rate of only $\sim 10^{-9}-10^{-8}$ yr$^{-1}$ in
the Galaxy \citep{kn88}. We instead suggest that part of these
ULXs may be results of the mergers of compact star binaries (e.g.,
neutron star-neutron star binaries, black hole-white dwarf
binaries, black hole-neutron star binaries). Here the accretion
disks originate from the disrupted, less massive stars. But the
accretion processes are likely to be similar as in the collapsar
model - after the initial hypercritical accretion (probably also
giving rise to GRBs), the accretion rate decreases to be around
the Eddington accretion rate, and the accreting compact stars
(most probably black holes) eventually appear as ULXs. The event
rate of these mergers may be high enough \citep{npp92,fry99} to
account for pop II ULXs.

\acknowledgements I am grateful to the anonymous referee for
insightful comments. This work was supported by NSFC and NKBRSF.


\begin{thebibliography}{}

\bibitem[Abramowicz et al.(1988)]{ab88}Abramowicz, M. A., Caerny, B. Lasota, J. P.,
\& Szuszkiewicz, E. 1988, ApJ, 332, 646
\bibitem[Balberg, Zampieri, \& Shapiro(2000)]{bzs00}Balberg, S., Zampieri, L.,
\& Shapiro, S. L., 2000, ApJ, 541, 860
\bibitem[Begelman(2002)]{b02}Begelman, M. C. 2002, ApJ, 568, L97
\bibitem[Bhattacharya \& van den Heuvel(1991)]{bh91}Bhattacharya, D.
\& van den Heuvel, E. P. J. 1991, Phys. Rep., 203, 1
\bibitem[Blair \& Fesen(1994)]{bf94}Blair, W. P.  \& Fesen, R. A. 1994, ApJ, 424, L103
\bibitem[Bloom et al.(1999)]{b99}Bloom, J. S. et al. 1999, Nature, 401, 453
\bibitem[Bodenheimer \& Woosley(1983)]{bw83}Bodenheimer, P. \&
Woosley, S. E. 1983, ApJ, 269, 281
\bibitem[Cannizzo, Lee, \& Goodman(1990)]{clg90} Cannizzo, J. K., Lee, H.
M., \& Goodman, J. 1990, ApJ, 351, 38
\bibitem[Chevalier(1989)]{c89}Chevalier, R. A. 1989, ApJ, 346, 847
\bibitem[Colbert \& Mushotzky(1999)]{cm99}Colbert, E. J. M. \& Mushotzky, R. F. 1999, ApJ, 519, 89
\bibitem[Colbert et al.(2003)]{c03}Colbert, E. J. M. et al. 2003, ApJ, submitted (astro-ph/0305476)
\bibitem[Fabbiano(1989)]{f89}Fabbiano, G., 1989, ARA\&A, 27, 87
\bibitem[Fabbiano et al.(2001)]{f01}Fabbiano, G. et al. 2001, ApJ, 554, 1035
\bibitem[Fruchter et al.(1999)]{f99}Fruchter, A. S. et al. 1999, ApJ, 519, L13
\bibitem[Fryer(1999)]{fry99}Fryer, C. L. et al. 1999, ApJ, 520, 650
\bibitem[Galama et al.(1998)]{g98}Galama, T. J. et al. 1998, Nature, 385, 670
\bibitem[Garnivich et al.(2001)]{g03}Garnivich, P. et al. 2003, ApJ, 582, 924
\bibitem[Hansen(1999)]{h99}Hansen, B. M. S. 1999, ApJ, 512, L117
\bibitem[Hjorth et al.(2003)]{h03}Hjorth, J. et al. 2003, Nature, 423, 347
\bibitem[Kaaret et al.(2003)]{k03}Kaaret, P., Corbel, S., Prestwich, A. H.,
\& Zezas, A. 2003, Science, 299, 365
\bibitem[King(2002)]{k02}King, A. R. 2002, MNRAS, 335, L13
\bibitem[King(2001)]{k01}King, A. R. et al. 2001, ApJ, 552, L109
\bibitem[Kohri \& Mineshige(2002)]{km02}Kohri, K. \& Mineshige, S.
2002, ApJ, 577, 311
\bibitem[K\"ording, Falcke, \& Markoff(2002)]{kfm02}K\"ording, E., Falcke, H.,
\& Markoff, S. 2002, A\&A, 382, L13
\bibitem[Kulkarni \& Narayan(1988)]{kn88}Kulkarni, S. R. \& Narayan, R. 1988, ApJ, 335, 755
\bibitem[Liu, Bregman, \& Seitzer(2002)]{lbs02} Liu, J. F., Bregman, J. N.,
\& Seitzer, P. 2002, ApJ, 580, L31
\bibitem[MacFadyen \& Woosley(1999)]{mw99} MacFadyen, A. I. \& Woosley, S. E. 1999, ApJ,
524, 262
\bibitem[Makishima et al.(2000)]{m00}Makishima, K. et al. 2000, ApJ, 535, 632
\bibitem[Meyer \& Meyer-Hofmeister(1989)]{mm89}Meyer, F. \& Meyer-Hofmeister, E. 1989,
in Thoery of Accretion Disks, ed. F. Meyer, W. Duschl, J. Frank,
\& E. Meyer-Hofmeister (Dordrecht: Kluwer), 307
\bibitem[Miller \& Hamilton(2002)]{mh02}Miller, R. C. \& Hamilton, D. P. 2002, MNRAS, 330, 232
\bibitem[Mineshige et al.(2002)]{m02}Mineshige, S.,  Hosokawa, T., Machida, M., \&
Matsumoto, R. 2002, PASJ, 54, 655
\bibitem[Mineshige, Nomoto, \& Shigeyama(1993)]{mns93}Mineshige, S., Nomoto, K.,
\& Shigeyama, T. 1993, A\&A, 267, 95
\bibitem[Mineshige et al.(1997)]{min97}Mineshige, S., Nomura, H.,
Hirose, M., Nomoto, K., \& Suzuki, T. 1997, ApJ, 489, 227
\bibitem[Narayan, Igumenshchev, \& Abramowicz(2000)]{n00}Narayan, R., Igumenshchev, I. V., \&
Abramowicz, M. A. 2000, ApJ, 539, 798
\bibitem[Narayan, Paczynski, \& Piran(1992)]{npp92}Narayan, R., Paczynski, B.,
\& Piran, T. 1992, ApJ, 395, L83
\bibitem[Narayan, Piran, \& Kumar(2001)]{n01}Narayan, R., Piran,
T., \& Kumar, P. 2001, ApJ, 557, 949
\bibitem[Ohsuga et al.(2002)]{o02}Ohsuga, K., Mineshige, S., Mori,
M., \& Umemura, M. 2002, ApJ, 574, 315
\bibitem[Paczynski(1998)]{p98}Paczynski, R. J. 1998, ApJ, 494, L45
\bibitem[Pakull \& Mirioni(2002)]{pm02}Pakull, M. W. \& Mirioni, L. 2002,
in New Visions of the X-ray Universe in the XMM-Newton and Chandra
Era, astro-ph/0202488
\bibitem[Podsiadlowski, Rappaport, \& Han(2003)]{prh03}Podsiadlowski, P., Rappaport, S.,
\& Han, Z. 2003, MNRAS, 341, 385
\bibitem[Roberts \& Colbert(2003)]{rc03}Roberts, T. P. \& Colbert, E. J. M. 2003, MNRAS, 341, L49
\bibitem[Roberts et al.(2003)]{r03}Roberts, T. P., Goad, M. R. , Ward, M.
J., \& Warwick, R. S. 2003, MNRAS, 342, 709
\bibitem[Sansuichi, Yamada, \& Fukue(1993)]{s93}Sansuichi, K., Yamada, T. T., \& Fukue, J. 1994, PASJ, 45,
443
\bibitem[Sarazin, Irwin, \& Bregman(2001)]{sib01}Sarazin, C. L., Irwin, J. A.,
\& Bregman, J. N. 2001, ApJ, 556, 533
\bibitem[Shakura \& Sunyaev(1973)]{ss73}Shakura, N. I. \& Sunyaev, R. A. 1973, A\&A, 24,
337
\bibitem[Woosley(1993)]{w93}Woosley, S. E. 1993, ApJ, 405, 273
\bibitem[Zezas et al.(2002)]{z02}Zezas, A., Fabbiano, G., Rots, A.
H., \& Murray, S. S. 2002, ApJ, 577, 710

\end{thebibliography}
\end{document}